\documentstyle[12pt]{article}
\begin{document}
\title{Multiply Connected Universes}
\author{B.G. Sidharth\\
International Institute for Applicable Mathematics \& Information Sciences\\
B.M. Birla Science Centre, Adarsh Nagar, Hyderabad - 500 063,
India}
\date{}
\maketitle
\begin{abstract}
It is now generally believed that our observable universe is one amongst a very large number - may be $10^{500}$ - of parallel universes. Following the author's own model in this context, we argue that this conglomeration of universes defines a multiply connected super space.
\end{abstract}
\section{Introduction}
The Newtonian universe was one in which there was an absolute background space
in which the basic building blocks of the universe were situated--these were stars.
This view was a quantum jump from the earlier view, based on the Greek model in which
stars and other celestial objects were attached to transparent material spheres, which
prevented them from falling down.\\
When Einstein proposed his General Theory of Relativity some ninety years ago, the
accepted picture of the universe was one where all major constituents were stationary.
This had puzzled Einstein, because the gravitational pull of these constituents should
make the universe collapse. So Einstein introduced his famous cosmological constant,
essentially a repulsive force that would counterbalance the attractive gravitational
force.\\
Shortly thereafter there were two dramatic discoveries which completely altered
that picture. The first was due to Astronomer Edwin Hubble, who discovered that the
basic constituents or building blocks of the universe were not stars, but rather
huge conglomerations of stars called galaxies. The second discovery was the fact that these
galaxies are rushing away from each other--far from being static the universe was exploding.
There was no need for the counterbalancing cosmic repulsion any more and Einstein dismissed
this as his greatest blunder.\\
Over the next forty odd years, these observations evolved into the Big Bang theory, according
to which all the matter in the universe possibly some fifteen billion years ago, was
concentrated in a speck, at the birth of the universe, which was characterized by an
inconceivable explosion or bang. This lead to the matter being flung outwards, and that
is what keeps the galaxies rushing outwards even today. In the mid sixties confirmation
for the Big Bang model of the universe came from the detection of a cosmic footprint. The
energy of the initial Big Bang would today still be available in the form of cosmic microwaves,
which accidentally were discovered \cite{ruffini,oh,mwt}.\\
Over the next three decades and more, the Big Bang theory was refined further and further. A
n important question was, would the universe continue to expand for ever, though slowing down,
or would the expansion halt one day and the universe collapse back again. Much depended on
the material content or density of the universe. If there was enough matter, then the
expansion would halt and reverse. If not the universe would expand for ever. However
the observed material content of the universe appeared to be insufficient to halt the
expansion.\\
At the same time there were a few other intriguing observations.
For example the velocities along the radius of a galaxy, instead
of sharply falling off, flattened out. All this led astronomers to
invoke dark matter, that is undetected matter. This matter could
be in the form of black holes within galaxies, or brown dwarf
stars which were too faint to be detected, or even massive
neutrinos which were otherwise thought
to be massless. With dark matter thrown in, it appeared that the universe had sufficient
material content to halt, and even reverse the expansion. That is, the universe would expand
up to a point and then collapse.\\
There still were several subtler problems to be addressed. One was the famous horizon problem.
To put it simply the Big Bang was an uncontrolled or random event and so different parts of
the universe in different directions were disconnected in the earliest stage and so today
need not be the same, just as people in different parts of the world need not wear the
same type of dress. Observation however shows that the universe is by and large uniform,
like people in different countries wearing the same dress--that would not be possible
without some form of intercommunication which would violate Einstein's earlier Special
Theory of Relativity, according to which no signal can travel faster than light. The
next problem was, that according to Einstein, due to the material content in the universe,
space should be curved or roughly speaking bent, whereas the universe appears to be flat.
There were other problems as well. For example astronomers predicted that there should be
monopoles that is, simply put, either only North magnetic poles or only South magnetic
poles, unlike the North South combined magnetic poles we encounter. Such monopoles have
failed to show up.\\
Some of these problems were sought to be explained by what has
been called inflationary cosmology where by, early on, just after
the Big Bang the explosion was a super fast \cite{zee,newcos}.\\
What would happen in this case is, that different parts of the universe, which could not
be accessible by light, would now get connected. At the same time, the super fast expansion
in the initial stages would smoothen out any distortion or curvature effects in space,
leading to a flat universe and in the process also eliminate the monopoles.\\
One other feature that has been studied in detail over the past
few decades is that of structure formation in the universe. To put
it simply, why is the universe not a uniform spread of matter and
radiation? On the contrary it is very lumpy with planets, stars,
galaxies and so on, with a lot of space separating these objects.
This has been explained in terms of fluctuations in density, that
is accidentally more matter being present in a given region.
Gravitation would then draw in even more matter and so on. Such
fluctuations would also cause the cosmic background radiation to
be non uniform or anisotropic. Such fluctuations are in fact being
observed.\\
Everything seemed to have fallen into place. The universe appeared to be well behaved - at least, well understood.\\
However in 1997 a diametrically opposite approach was proposed by the author \cite{ijmpa,ijtp,mg8,uof}, in which matter is created at random from a background Quantum Vacuum or dark energy.Interestingly the
random nature of creation of the particles take place in a fashion similar to inflation,
though this effect would be pronounced in the earliest stages of the birth of the universe,
as in the inflation model. This model successfully predicted an ever expanding and accelerating
universe. It explains a number of mysterious relations between different physical and
astronomical quantities, for example the radius of the universe, the number of particles
in the universe, the mass and size of a typical elementary particle, the universal
gravitational constant, the speed of light and so on.\\
Many of these puzzling relations
had been known for a long time, but in the absence of an explanation, they had been
dismissed as freak coincidences. In the present model, all these relations follow from
the theory, rather than being accidental. Apart from the fact that this model provides
an explanation for the puzzling time variation of the fine structure constant, it also
gives a mechanism for reconciling the two great irreconcible theories of the twentieth
century, namely Einstein's General Relativity and Quantum Theory. The key to this is the
fact that, in both these theories, space and time were taken to be continuous and smooth,
whereas in this model this is no longer true, though these subtler effects can only be
detected at very very tiny scales or high energies.\\
However from early 1998 the conventional wisdom of cosmology that had developed over the
past three to four decades, began to be challenged \cite{perl,nature}. The work of Perlmutter and co-workers,
as also of Schmidt and co-workers was announced in 1998 and it told a sensational if
different story. They had observed carefully very distant supernovae or exploding stars
and to the disbelief of everyone, it turned out that not only was the universe not slowing
down, but was actually accelerating, which also means that it would keep on expanding
eternally. This was nothing short of an upheaval, and theorists are
under great pressure to explain all this. Of course, this confirmed the author's prediction.\\
Suddenly astronomers were talking about dark energy instead of dark matter. Dark energy is an
unknown and mysterious form of energy that brings into play repulsion, over and above the attractive force of gravitation. All this is reminiscent of Einstein's greatest blunder
namely the cosmic repulsion itself.\\
However there is a problem. What is dark energy? Physicists speak of such an energy from
what is called the Quantum Vacuum. The idea here is that there cannot be a background vacuum
with exactly zero energy as exact values of energy are forbidden by Quantum Theory. Only the
average energy could be zero. In other words energy would be fluctuating about a zero value.
This is called a Zero Point Field. As a consequence what happens in the vacuum is that
electrons and positrons are continuously created, out of nothing as it were, but these pairs
are very shortlived. Almost instantaneously they annihilate each other and release energy,
which in turn again manifests itself as electron-positron pairs. These effects could lead
to a cosmic repulsion, but the only problem is that the value of the cosmological constant,
in other words the strength of the cosmic repulsion would be much too high. This would go
against observation. The problem has been known for long as the cosmological constant problem \cite{weinberg,contphys}.
So astronomers are also talking about the dark energy, christened Quintessence by some, leading
to a mysterious new force of repulsion.\\
Yet another dramatic discovery since 1998 has been made with the
help of the SuperKamiokande experiment in Japan \cite{susy}. This
facility observed solar radiation, in particular for the very
strange, maverick supposedly massless particles, neutrinos. It
turns out  that these particles now possess a miniscule mass,
about a billionth that of an electron. The discovery explains one
puzzle, what has been commonly called the solar neutrino problem.
The point is that we seem to receive much less than the
theoretically expected number of neutrinos from solar radiation.
But the theoretical prediction was made on the basis of the
assumption that neutrinos were massless. Even with the tiniest of
masses, the problem disappears. However these observations
challenge what has come to be known as the Standard Model of
Particle Physics, which takes the masslessness of the neutrino for
granted. Could these massive neutrinos be the elusive dark matter?
The answer is no-- the mass of this matter is still much too small to stop the
expansion, which is very well in view of the
latest ever expanding scenario.\\
Another iconoclastic dramatic observation which is gaining
confirmation is that, what is called the fine structure constant,
which scientists have considered to be a sacrosanct constant of
the universe, has infact been slowly decreasing over billions of
years \cite{webb1,webb2}. Webb and co-workers have confirmed this by
observing the spectrum of light from the distant Quasars and
comparing this with spectra in the vicinity. As the fine structure
constant is made up of the electric charge of the electron, the
speed of light, and the Planck constant, this would mean that one
or some or even all of them are not the sacred constants they have
been taken for, but are slowly changing with time. The
consequences of this are quite dramatic. For instance this would
mean that atoms and molecules in the past were not the same as
their counterparts today - this will be true in future also. This
again would have dramatic implications. According to present
thinking, life as we know it depends on a delicate balance between
the different fundamental constants of nature--otherwise life
itself could neither evolve nor sustain. If the values of these
constants change, so would atoms and molecules and the narrow
limits for life get narrower in time.\\
While attempts are being made to modify the successful inflation
theory and other theories also to try to explain these latest
discoveries there are alternative approaches being considered. For example a theory that has been put forward
in the past few years by Moffat, Albrecht, Magueijo, Barrow and
others that contrary to Einstein's Special Relativity the speed of
light is not a universal constant, but rather it has decreased
over billions of years \cite{mof}.
\section{Different Routes to Multiple Universes}
As we saw astronomers believe that our universe was born in a titanic explosion of the skind that would put into shade trillions of trillions of Hydrogen bombs exploding at once. The out rushing matter then condensed into galaxies, stars, planets and other known - and who knows?- unknown objects over a period of fourteen billion years. During this time the universe has swollen to a size of some million, billion, billion kilometers. This distance defines a horizon beyond which even our most powerful telescopes are unable to peer.\\
But is this horizon the boundary where the story ends? Probably not, is the view to which many astronomers are veering. Beyond the horizon of our direct perception, the universe extends to infinity, to any number of trillions of such universes strewn all across. And who knows, there would be any number of planets identical, or almost identical to our own earth, complete with human beings like us. Even with a carbon copy of yourself. In fact we can do better. Simple calculations reveal that there would be a galaxy identical to ours, one followed by the number of digits in the radius of our universe expressed in kilometers--that is twenty five--that many kilometers away. All the different universes are parallel universes. Multiverse is the name going around these days.\\
But even this is not the end of the story. According to other ideas, such a conglomeration of  parallel universes is what may be called the Level One Multiverse, something that can be thought of as a bubble. According to modern theories,  inflation or the super fast blow out could well have created any number of such bubbles, each bubble being a huge collection of parallel universes. One could think of these bubbles or the Level One Multiverses as the many individual bubbles gushing out of a bottle of an aerated drink that has just been opened.\\
What could distinguish the various Level One Multiverses from each other? In fact they could all be very different with very different values of the electric charge, the gravitational constant and what not.\\
The story does not stop even here! There could be Multiverses here and now! That would be the case if an interpretation of the century old Quantum Theory due to Hugh Everett 3 is correct \cite{everett}. In fact many physicists think it is. According to our usual physics, a particle travelling from a point A to a point B is a fairly straightforward event. Quantum Theory however, tells us that the particle could go from A to B in many different ways other than what we see. The usual interpretation is that by the very act of observing the particle more from A to B we eliminate all other possibilities. This has been the generally accepted point of view. But in 1957 Everett proposed that all other possibilities which have supposedly been snuffed out actually do take place, and its only one of them which we get to see. So in the simple act of, let us say an electron going from one point to another point, a centimeter away, there are millions of hidden acts that have taken place, each in its universe.\\
And this is not all. Einstein's Theory of Relativity leads us to the conclusion that when a star collapses into a black hole, at the very center there is what is called a singularity. A singularity can be thought of as a junction or crossroads of infinitely many different roads. Only in our case it is the junction of infinitely many universes each legitimate in its own right, and moreover each with its own laws of nature. This is because at the singularity itself, all laws of nature breakdown, just as, exactly at the North Pole there is no meaning to the East, West or other directions.\\
Some scientists have also come to a similar conclusion from the view point of Superstring theory. It is reckoned that there would be so many universes … well, one followed by some five hundred zeros! \cite{phys}\\
The author's own theory of the universe of universes shares some of the above features. It turns out that the universe can be thought to be a blown up version of an elementary particle which is spinning. A huge number of such "particle universes" would form the analogue of a super particle universe--the analogue of let us say, gas molecules in a cubic centimeter. And so on, possibly with increasing dimensionality of space and time at each step \cite{sss}. It is a bit like colonies of colonies of colonies and so on. Ofcourse all this is not a fantasy dreamt out by scientists. There are very delicate tests proposed which can provide a clue.
\section{Multiply Connected Universes}
We first consider descriptions from microphysics on which we can model multiple universes. Let us consider the usual Hamilton-Jacobi theory of Classical Mechanics \cite{gold}. Here we start with the action integral
\begin{equation}
I = \int L \left(\frac{dx}{dt}, x, t\right) dt\label{e1}
\end{equation}
and extremalize it \cite{wheeler,mwt}. Let us denote, working for the moment in one space dimension for simplicity, the extremum integral of (\ref{e1}) by
\begin{equation}
S (x,t) = \bar{I}\label{e2}
\end{equation}
We then have,
\begin{equation}
p = \frac{\partial S}{\partial x}\label{e3}
\end{equation}
\begin{equation}
E = - \frac{\partial S}{\partial t} = \frac{p^2}{2m} + V (x)\label{e4}
\end{equation}
where $p$ denotes the momentum and $E$ denotes the energy. We can combine (\ref{e3}) and (\ref{e4}) and write,
\begin{equation}
E = H (p,x)\label{e5}
\end{equation}
Equation (\ref{e5}) is well known in Classical Mechanics but let us now introduce a description in terms of the wave function
\begin{equation}
\psi = Re^{(\imath/\hbar )S(x,t)}\label{e6}
\end{equation}
where $R$ is slowly varying and $S$ is given by (\ref{e2}). We now get from (\ref{e6}) the Hamilton-Jacobi equation in one dimension
\begin{equation}
- \frac{\partial S}{\partial t} = H \left(\frac{\partial S}{\partial x} , x\right) = \frac{1}{2m} \left(\frac{\partial S}{\partial x}\right)^2 + V (x)\label{e7}
\end{equation}
whose solution is
\begin{equation}
S = (x,t) = - Et + S \left\{ 2m\left[E - Vjk\right]\right\}^{1/2} dx + S\label{e8}
\end{equation}
In view of the description given in (\ref{e6}), (\ref{e7}) and (\ref{e8}) take on a more general character in terms of the so called dynamical phase $S$ . Further as is well known, the requirement 
$$\frac{\partial S}{\partial E} = 0$$
which denotes a constructive interference in the phase of systems described by wave functions leads us back to a classical particle description as in (\ref{e4}). We will now apply the above theory to Quantum Mechanical systems and General Relativity.\\
In Quantum Theory, the description of the wave function as in (\ref{e6}) when substituted in the Schrodinger equation 
\begin{equation}
\imath \hbar \frac{\partial \psi}{\partial t} = - \frac{\hbar^2}{2m} \nabla^2 \psi + V \psi\label{e9}
\end{equation}
leads to
\begin{equation}
\frac{\partial \rho}{\partial t} + \vec{\nabla} \cdot (\rho \vec{v}) = 0\label{e10}
\end{equation} 
\begin{equation}
\frac{1}{\hbar} \frac{\partial S}{\partial t} + \frac{1}{2m} (\vec{\nabla}S)^2 + \frac{V}{\hbar^2} - \frac{1}{2m} \frac{\nabla^2 R}{R} = 0\label{e11}
\end{equation}
where $\rho = R^2,\vec{v} = \frac{\hbar}{m} \vec{\nabla} S \, \mbox{and}\, Q \equiv - \frac{\hbar^2}{2m} (\nabla^2 R/R)$.\\ Equation (\ref{e10}) is easily recognized as the equation of continuity and (\ref{e11}) as the Hamilton-Jacobi equation in three dimensions.\\
Let us now look at Quantum Theory from a slightly different point of view. We start by reviewing Dirac's original derivation of the Monopole \cite{dirac,dirac2}.  He started with the wave function
\begin{equation}
\psi = Ae^{\imath \gamma},\label{e1a}
\end{equation}
He then considered the case where the phase $\gamma$ in (\ref{e1}) is non integrable. In this case (\ref{e1a}) can be rewritten as
\begin{equation}
\psi = \psi_1 e^{\imath S},\label{e2a}
\end{equation}
where $\psi_1$ is an ordinary wave function with integrable phase, and further, while the phase $S$ does not have a definite value at each point, its four gradient viz., 
\begin{equation}
K^\mu = \partial^\mu S\label{e3a}
\end{equation}
is well defined. We use temporarily natural units, $\hbar = c = 1$. Dirac then goes on to identify $K$ in (\ref{e3a}) (except for the numerical factor $hc/e$) with the electromagnetic field potential, as in the Weyl gauge invariant theory.\\
Next Dirac considered the case of a nodal singularity, which is closely related to what was later called a quantized vortex (Cf. for example ref.\cite{bgsmonopole}). In this case a circuit integral of a vector as in (\ref{e3a}) gives, in addition to the electromagnetic term, a term like $2 \pi n$, so that we have for a change in phase for a small closed curve around this nodal singularity,
\begin{equation}
2 \pi n + e \int \vec B \cdot d \vec S\label{e4a}
\end{equation}
In (\ref{e4a}) $\vec B$ is the magnetic flux across a surface element $d \vec S$ and $n$ is the number of nodes within the circuit. The expression (\ref{e4a}) directly lead to the Monopole in Dirac's formulation.\\
Let us now reconsider the above arguments in terms of recent developments. The Dirac equation for a spin half particle throws up a complex or non Hermitian position coordinate \cite{diracpqm,bgsfpl}. Dirac identified the imaginary part with zitterbewegung effects and argued that this would be eliminated once it is realized that in Quantum Mechanics, spacetime points are not meaningful and that on the contrary averages over intervals of the order of the Compton scale have to be taken to recover meaningful physics. Over the decades the significance of such cut off space time intervals has been stressed by T.D. Lee and several other scholars as noted earlier \cite{lee,bom,kardy,ppp,psu}. Indeed with a minimum cut off length $l$, it was shown by Snyder that there would be a non commutative but Lorentz invariant spacetime structure. At the Compton scale we would have \cite{snyder},
\begin{equation}
[x,y] = 0(l^2)\label{e5a}
\end{equation}
and similar relations.\\
In fact starting from the Dirac equation itself, we can deduce directly the non commutativity (\ref{e5a}) (Cf.refs.\cite{bgsfpl}).\\
Let us now return to Dirac's formulation of the monopole in the light of the above comments. As noted above, the non integrability of the phase $S$ in (\ref{e2a}) gives rise to the electromagnetic field, while the nodal singularity gives rise to a term which is an integral multiple of $2 \pi$. As is well known we have
\begin{equation}
\vec \nabla S = \vec p\label{e6a}
\end{equation}
where $\vec p$ is the momentum vector. When there is a nodal singularity, as noted above, the integral over a closed circuit of $\vec p$ does not vanish. In fact in this case we have a circulation given by
\begin{equation}
\Gamma = \oint \vec \nabla S \cdot d \vec r = \hbar \oint dS = 2 \pi n\label{e7a}
\end{equation}
It is because of the nodal singularity that though the $\vec p$ field is irrotational, there is a vortex - the singularity at the central point associated with the vortex makes the region multiply connected, or alternatively, in this region we cannot shrink a closed smooth curve about the point to that point. In fact if we use the fact as seen above that the Compton wavelength is a minimum cut off, then we get from (\ref{e7a}) using (\ref{e6a}), and on taking $n = 1$,
\begin{equation}
\oint \vec \nabla S \cdot d\vec r = \int \vec p \cdot d \vec r = 2\pi mc \frac{1}{2mc} = \frac{h}{2}\label{e8a}
\end{equation}
$(l = \frac{\hbar}{2mc}$ is the radius of the circuit and $\hbar = 1$ in the above natural units). In other words the nodal singularity or quantized vortex gives us the mysterious Quantum Mechanical spin half (and other higher spins for other values of $n$). In the case of the Quantum Mechanical spin, there are $2 \times n/2 + 1 = n + 1$ multiply connected regions, exactly as in the case of nodal singularities.\\
Indeed we can see the role of a multiply connected space even within the context of the non relativistic Schrodinger equation. We consider a Hydrogen like atom in two dimensional space, for which the Schrodinger equation is given by \cite{ho,bgsarxiv}
\begin{equation}
-\frac{\hbar^2}{2\mu} \left[\frac{1}{r} \frac{\partial}{\partial r} \left(r\frac{\partial}{\partial r}\right) + \frac{1}{r^2}\frac{\partial^2}{\partial \phi^2}\right] \psi(r,\phi) - \frac{Ze^2}{r}\phi(r,\phi) = E\psi(r,\phi)\label{e13a}
\end{equation}
As is well known the energy spectrum for (\ref{e13a}) is given by
\begin{equation}
E = -\frac{Z^2e^4\mu}{2\hbar^2(n+m+\frac{1}{2})^2}\label{e14a}
\end{equation}
If we require that (\ref{e14a}) be identical to the Bohr spectrum, then $m$ should be a half integer, which also means that the configuration space is multiply connected. In the simplest case of a doubly connected space, we are dealing wityh $R^2 \times S^1$, where $S^1$ is a compactified space, generally considered to be a Kaluza-Klein space.\\
With the above background let us return to the universe at large. We consider Wheeler's description of the wave function of the universe in terms of super space \cite{wsp}. Superspace is a manofold such that a single point therein represents the whole of the geometry of three dimensional space. As is usual we denote such a three dimensional geometry by (\ref{e3}) $G$ . Treating this as a point we introduce the wave function like description (\ref{e6}), except that we are in superspace which is essentially a four dimensional manifold.\\
Interestingly it has been shown by Stern that as long as we are dealing with Euclidean type three dimensional spaces with positive definite metric, superspace constitutes a manifold in the sense that each point therein has a neighborhood homeomorphic to an open set in a Banach space and two distinct points are separated by disjoint neighborhoods. This enables us to carry out the usual operations in superspace and we are led back using the principle of constructive interference as earlier, to this time the ten field equations of  Einstein.\\
Thus the Hamilton-Jacobi theory leads both to Quantum Mechanics and to General Relativity in terms of the wave function description (\ref{e6}) together with constructive interference. However it must be borne in mind that in Quantum Mechanics we are dealing with the usual three dimensional space whereas for obtaining the Einstein's equations of General Relativity, we are using the four dimensional superspace.\\ 
To push these considerations further let us consider the general relativistic formulation in terms of linearized theory. In fact in the case of the electron,  it was shown \cite{cu} that the spin was given by, 
\begin{equation}
S_K = \int \epsilon_{klm} x^lT^{m0} d^3 x = \frac{h}{2}\label{eb1}
\end{equation}
where the domain of integration was a sphere of radius given by the Compton wavelength. If this is carried over to the case of the universe, we get from (\ref{eb1})
\begin{equation}
S_U = N^{3/2} h \approx h_1\label{eb2}
\end{equation}
where $h_1$ and $S_U$ denotes the counterpart of electron spin (Cf.ref.\cite{cu}).\\
$h_1$ in (\ref{eb2}) turns out to be the spin of the universe itself in broad agreement with Godel's spin value for Einstein's equations \cite{godel,carneiro}. Incidentally this is also in agreement with the Kerr limit of the spin of the rotating Black Hole. Further as pointed out by Kogut and others, the angular momentum of the universe given in (\ref{eb2}) is compatible with a rotation from the cosmic background radiation anisotropy \cite{carneiro}. Finally it is also close to the observed rotation as deduced from anisotropy of cosmic electromagnetic radiation as reported by Nodland and Ralston and others \cite{ralston,kogut}.\\
In the above $h_1 \sim 10^{93}$ and we immediately have
\begin{equation}
R = \frac{h_1}{Mc}\label{eb3}
\end{equation}
where $R$ the radius of the universe is the analogue of the particle Compton wavelength in the macro context and $M$ is the mass.\\
Thus the multiply connected space considerations of micro physics which we saw above can immediately be taken over to the case of the macro cosmos. In this case super space would be multiply connected consisting of three (space) dimensional universes which mimick particles with spin.

\end{document}